\documentclass[aps,pre,twocolumn]{revtex4}
\usepackage{graphicx}
\usepackage{epsfig}
\usepackage{amsmath}
\usepackage{graphics}
\usepackage{epsfig}
\usepackage{tikz}
\newcommand{\bra}[1]{\ensuremath{\langle{#1}|\,}}
\newcommand{\ket}[1]{\ensuremath{\,|{#1}\rangle}}

\begin{document}

\title{Distribution of tunnelling times for quantum electron transport}


\author{Samuel L. Rudge}
\author{Daniel S. Kosov}
\address{College of Science, Technology and Engineering, James Cook University, Townsville, QLD, 4811, Australia }
\url{http://www.hbar-electronics.info}

\pacs{05.30.-d, 05.60.Gg, 72.10.Bg}

\begin{abstract}
In electron transport, the tunnelling time is the time taken for an electron to tunnel out of a system after it has tunnelled in.
We define the tunnelling time distribution for quantum processes in a dissipative environment and develop a practical approach for calculating it, where the environment is described by the general Markovian master equation. 
We illustrate the theory by using the rate equation to compute the tunnelling time distribution for electron transport through a molecular junction. The tunnelling time distribution 
is exponential, which indicates that Markovian quantum tunnelling is a Poissonian statistical process.  The tunnelling time distribution is used not only to study the quantum statistics of tunnelling along the average electric current but also to analyse extreme quantum events where an
electron jumps against the applied voltage bias. The average tunnelling time shows distinctly different temperature dependance for p- and n-type molecular junction and therefore provides a sensitive tool to probe the alignment of molecular orbitals relative to the electrode Fermi energy.
\end{abstract}

\maketitle

\section{introduction}

Single molecule electronics is an exciting new area of science with a broad range of important applications such as in energy harvesters, chemical sensors, and new elements of computer hardware \cite{moletronics}.  Single molecule junctions have also been developed into an integral part of modern chemical physics, where complex fundamental quantum mechanical questions can be studied theoretically and experimentally \cite{Aradhya:2013aa}.

 Electric current is obtained by averaging a macroscopic number of transferred electrons over a large  time interval, which greatly diminishes the significant amount of interesting physical information obtained.  Studies beyond average electric current, such as noise, full counting statistics, and fluctuation relations are becoming more common  both experimentally and theoretically 
 \cite{PhysRevB.90.075409,PhysRevB.92.245418,PhysRevB.84.205450,PhysRevB.87.115407,PhysRevB.91.235413,avriller09,thoss14,segal15}.
 Noisel measurements have recently been used to characterise microscopic details of electron transport through molecules\cite{PhysRevLett.100.196804,doi:10.1021/nl201327c,doi:10.1021/nl060116e,Tsutsui:2010aa}.
  
 Electron transport through a molecule can be considered as a series of electron quantum tunnelling events that occur at specific but random times. Such stochastic processes  are often statistically described in terms of a waiting time distribution \cite{vanKampen}. 
Waiting time distributions between quantum events were
first introduced in quantum mechanics by Srinivas and Davies
in 1981 to describe photon counting experiments \cite{Srinivas2010}.
Their work was an extension of early work by Mollow et al. in quantum statistics \cite{Scully1969,Mollow1968}. While
photon counting experiments were traditionally treated classically, a quantum
mechanical description is needed to incorporate the continuous measurements
of photodetectors. The theoretical work on waiting time distributions for
photon emissions led to applications in characterising quantum jumps
in atomic systems \cite{Zoller1987}.

The use of waiting time distributions to study electron transport in nanoscale systems was pioneered by
Brandes \cite{brandes08}.
Brandes presented the methodology to derive the 
waiting time distribution for electron transport through an open quantum system described by a Markovian master equation  \cite{brandes08}. Additionally,
he showed that waiting time distributions  can be used to calculate shot noise, various current
fluctuations, and is related to full counting statistics  \cite{brandes08}.
His version of the waiting time formalism forms the basis of our
discussion in the following sections.

In this paper we propose a  physical quantity - the tunnelling time distribution - which provides new physical information and is sensitive to details of electron transport processes. The tunnelling time distribution is  a ``spin-off" quantity  from the 
waiting time distribution between photon detections (discussed in quantum optics since early 80s) and the waiting time distribution between electric current spikes \cite{brandes08,buttiker12}.
The first moment of the tunnelling time distribution is the average tunnelling time, which is the average time that an electron spends in the molecule before it is transferred into the source electrode.
The unambiguous definition of the average tunnelling time has been a much debated topic in physics \cite{RevModPhys.66.217,RevModPhys.61.917}.
Knowledge of the average tunnelling time gives useful insight on various chemical processes, such as nonadibatic electron transfer \cite{nitzan00} and the breakdown of the Born-Oppenheimer approximation \cite{doi:10.1021/acs.jpcb.5b00862}.

The paper is organised as follows. Section II describes the general derivations of the tunnelling time distribution for Markovian quantum events.  
In section III, we apply the theory to electron transport through molecular junctions and physically interpret the obtained results.
Section IV summarises the main results of the paper. Appendix A contains technical details and proofs of mathematical identities used in the paper.

We use natural units in equations throughout the paper:
$\hbar = k_B = |e| = 1$, where $-|e|$ is the electron charge.

\section{Distribution of tunnelling time for Markovian quantum dynamics}

Quantum transport of electrons through a  molecular junction is a stochastic process that consists of  a  series of  quantum ``jumps"  (electron tunnelling events) that take place at specific but random times. Suppose that we observe an electron tunnelling from the source electrode into the molecule at an initial time $t_1$, and then
 we observe an electron tunnelling out of the molecule into the drain electrode at some later  time  $t_2$.
The times of the electron tunnelling  events, $t_1$ and $t_2$,  are stochastic variables distributed on the positive real axis
\begin{equation}
0\le t_1 \le t_2 \le  + \infty.
\end{equation}

Let us define the following probability distributions:
\\
$P(t_2, t_1)$  -- the joint probability distribution that an electron 
 tunnels into the molecule  at time  $t_1$
 and tunnels out of the molecule at time   $t_2$,
\\
$p(t)$  -- the probability distribution that an electron 
 tunnels from the source electrode into the molecule  in time  $t$,
 \\
 $w(t_2, t_1)$ -- the conditional probability that  an electron tunnels to the drain electrode at time $t_2$ given that it was transferred to the molecule from the source electrode at time $t_1$ and {\it that there were no other tunnelling events between time $t_1$ and $t_2$} \footnote{This depends on the splitting of the Liouvillian. In this paper we consider the case where the Liouvillian is separated so that we monitor all quantum jumps}.
 We will call  $w(t_{2},t_{2})$ the {\it tunnelling time distribution}.
 
In the  nonequilibrium steady state regime as well as in thermodynamic equilibrium, $w(t_2, t_1)$  and  $P(t_2, t_1)$ depend on relative time only, $\tau = t_2 -t_1$, and $p(t)$ becomes time-independent.

These three probability distributions are related to each other by the standard Kolmogorov relation between joint and conditional probabilities,
which is physically interpreted as:
\begin{equation}
P(t_2, t_1) =  w(t_2, t_1) p(t_1).
\label{kolmogorov1}
\end{equation}
In the steady state regime, (\ref{kolmogorov1}) becomes
\begin{equation}
P(\tau) =  w(\tau) p.
\label{kolmogorov2}
\end{equation}
Our goal is to derive the explicit expression for the tunnelling time distribution based on the master equation description of electron transport 
through a quantum system.

\begin{figure}
\begin{tikzpicture}
\shade[bottom color=black, top color=white] (-0.5,0) rectangle (1.5,3.2);
\draw (-0.5,0) rectangle (1.5,4) node[above= 2cm, midway] {       source electrode};
\draw [black, thick, domain=-0.5:1.5] plot (\x, {2.1+(\x-0.5)*(\x-0.5)*(\x-0.5)}); 
\draw[line width = 0.5pt, dashed] (-0.5,2.1) -- (1.5,2.1) node[midway, above] {$\mu_L$};
%

\shade[bottom color=black, top color=white] (6,0) rectangle (8,2);
\draw (6,0) rectangle (8,4) node[above= 2cm, midway] { drain electrode};
\draw [black, thick, domain=6:8] plot (\x, {1+(\x-7)*(\x-7)*(\x-7)}); 
\draw[line width = 0.5pt, dashed] (6,1) -- (8,1) node[midway, above] {$\mu_R$};
%

\draw[line width = 2pt] (3.3,1.6) -- (4.2,1.6) node[midway, below] {$\epsilon$};
\draw (3.75,1.6) circle [radius=1] node[above=1cm] {molecule};

\draw [ ->, green,  very thick ] (1.,1.65) arc(150:30:40pt) node[midway, above] {$J_{\text{sm}}$};
\draw [ <-, red, very thick ] (-0.45,1.55) arc(-150:-30:64.5pt)  node[pos=0.57, below] {$J_{\text{ms}}$};;

\draw [ ->, green,  very thick ] (4.1,1.65) arc(150:30:40pt) node[midway, above] {$J_{\text{md}}$};
\draw [ <-, red, very thick ] (4,1.55) arc(-150:-30:64.5pt)  node[pos=0.44, below] {$J_{\text{dm}}$};;

\end{tikzpicture}
\caption{Sketch of the model molecular junction and quantum jumps superoperators resposible for electron transport. The quantum jumps superoperators are given by (\ref{jsm}, \ref{jmd}, \ref{jms}, \ref{jdm}) and collectively they represent dissipative part of the Liouvillian (\ref{L}). Green arrows show quantum jumps along the average current flow, red arrows depict quantum tunnelling agains voltage bias. }
\label{sketch}
\end{figure}
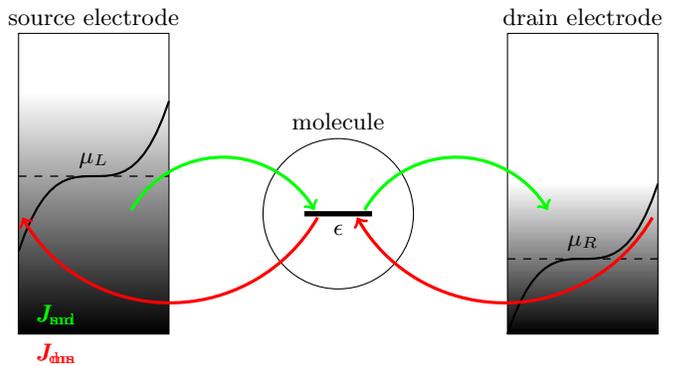

Let us consider a general Markovian master equation
\begin{eqnarray}
\dot \rho(t) =  {L} \rho(t).
\label{liouville}
\end{eqnarray}
Here $  {L}$ is a  Liovillian superoperator that contains both the Hermitian part and the dissipative non-Hermitian superoperators which originate from molecule-electrode coupling.
The Liouvillian can be formally split into two parts: the ``jumpless" part $ {L}_{0}$
and a part that describes the quantum jump electron tunnelling processes. There are four generic tunnelling processes in single electron transport, which are illustrated in Fig.\ref{sketch}: $ {J}_{\text{sm}}$ -- tunnelling from the source electrode into the molecule, $ {J}_{\text{ms}}$ - tunnelling from the molecule into the source electrode 
$ {J}_{\text{md}}$ -- tunnelling from the molecule into the drain electrode, and $ {J}_{\text{md}}$ -- tunnelling from the drain electrode into the molecule:
\begin{eqnarray}
\label{L}
 {L} & = &  {L}_{0}+ {J}_{\text{sm}}  +  {J}_{\text{ms}} +  {J}_{\text{dm}}  +  {J}_{\text{md}}.
\end{eqnarray}
If we use molecular eigenstates for the spectral decomposition of the Liovillian, then the quantum jump operators always occur on off-diagonals of the Liouvillian super-matrix and
the $ {J}$ superoperators have zero diagonal elements.
The  particular examples of how to extract quantum jump super-operators from the Liouvillian  will be discussed in later sections. However, it is pertinent
to note that the splitting is not unique as it depends on the type
and number of quantum jump operations monitored.

To develop the formalism associated with the tunnelling time distribution, we let the system evolve to the nonequilibrium steady state -- it is described by the steady state density matrix, which is the null vector of the full Liouvillian --
\begin{equation}
  {L} \; \overline \rho =0,
\end{equation}
and then we begin to monitor tunnelling quantum jumps.
First, we define the joint probability distribution that an electron 
 tunnels into the molecule from the source electrode in some arbitrary time  $t$
 and `jumps' (tunnels) out of the molecule to the drain electrode in time   $t +\tau$:
\begin{eqnarray}
P(\tau) = \text{Tr} [ {J}_{\text{md}}e^{ {L}_{0}\tau} {J}_{\text{sm}} \overline\rho].
\label{P}
\end{eqnarray}
This definition is self-explanatory: the system is prepared in state $\overline\rho$, it undergoes quantum jump $J_{\text{sm}}$ (tunnelling of an electron from the source electrode), then the system evolves without experiencing any of the monitored tunnelling events  for time $\tau$ and finally it undergoes the quantum jump $J_{\text{md}}$ (tunnelling of an electron to the drain electrode). Since the probability that an electron tunnels into the molecule in arbitrary time after the establishment of the steady state is
\begin{equation}
p = \text{Tr} [ {J}_{\text{sm}} \overline\rho],
\label{p}
\end{equation}
we rewrite (\ref{P}) as
\begin{equation}
P(\tau) = \frac{\text{Tr} [  {J}_{\text{md}}e^{ {L}_{0}\tau} {J}_{\text{sm}} \overline\rho]}{\text{Tr} [ {J}_{\text{sm}} \overline\rho]} \;  \text{Tr} [ {J}_{\text{sm}} \overline\rho].
\label{P2}
\end{equation}
Comparing (\ref{p})  and (\ref{P2}) with (\ref{kolmogorov2}), we identify the expression for the tunnelling time distribution 
\begin{equation}
w(\tau) = \frac{\text{Tr} [  {J}_{\text{md}}e^{ {L}_{0}\tau} {J}_{\text{sm}} \overline\rho]}{\text{Tr} [ {J}_{\text{sm}} \overline\rho]}. 
\label{w}
\end{equation}
The tunnelling time distribution is not normalised, as it does not contain all the probability for all secondary quantum tunnelling events. We expect that over all time $[0,\infty)$ the probability for a quantum tunnelling event to occur is unity. However, since we are monitoring four different quantum tunnelling  events (described by superoperators ${J_\alpha}$, where $\alpha={\text{sm},\text{ms},\text{dm},\text{md}}$), we must sum over all possible secondary quantum jumps $\sum_{\alpha}$ and integrate over time $\tau$ for normalisation:
\begin{eqnarray}
  \sum_{\alpha} \int_0^{\infty} d \tau  \frac{\text{Tr} [  {J}_{\alpha}e^{ {L}_{0}\tau} {J}_{\text{sm}} \overline\rho]}{\text{Tr} [ {J}_{\text{sm}} \overline\rho]} && 
  \label{norm}
  \\
  = \int_0^{\infty} d \tau  \frac{\text{Tr} [ \sum_{\alpha}   {J}_{\alpha}e^{ {L}_{0}\tau} {J}_{\text{sm}} \overline\rho]}{\text{Tr} [ {J}_{\text{sm}} \overline\rho]}  
 &=&-  \frac{\text{Tr} [ ( {L} -  {L}_{0})  {L}_{0}^{-1}  {J}_{\text{sm}} \overline\rho]}{\text{Tr} [ {J}_{\text{sm}} \overline\rho]}  
\nonumber
\end{eqnarray}
Since $\text{Tr}[ {L} A]=0$ (where $A$ is an arbitrary matrix), normalisation (\ref{norm}) becomes
\begin{equation}
 \frac{\text{Tr} [  {L}_{0}  {L}_{0}^{-1}  {J}_{\text{sm}} \overline\rho]}{\text{Tr} [ {J}_{\text{sm}} \overline\rho]} =  \frac{\text{Tr} [   {J}_{\text{sm}} \overline\rho]}{\text{Tr} [ {J}_{\text{sm}} \overline\rho]}  =1.
\end{equation}
%
%
From here we further simplify the tunnelling time distribution expression and bring it to a computationally convenient form by applying super-operator notation.\cite{brandes08,harbola08,dzhioev11a, dzhioev11b,dzhioev12,dzhioev14,dzhioev15}
Here we will closely follow the derivations for the waiting time distribution proposed by  Brandes \cite{brandes08}.

Linear operators in the Hilbert  space form themselves a linear  vector space:
for each linear operator $A$ we put in correspondence a
  (super) vector $|A)$.
The  scalar product  is defined as  $(A_1|A_2) = \mathrm{Tr}(A_1^\dag A_2)$, where $A_1$ and $A_2$ are two linear operators.
In particular, the scalar product of a vector corresponding to some operator $A$ and a vector corresponding to  the identity operator $I$ 
is equivalent to the trace operation in the Fock space: $(I|A)=\mathrm{Tr}(IA)= \mathrm{Tr}(A)$.
If $\rho(t)$ is the density matrix of the system, then $|\rho(t) )$ is the supervector and $(I| \rho(t) ) =1$.

The operators acting on super-vectors, such as Liouvillians $  L$, $  L_0$ and quantum jumps $ {J}_{\alpha}$ , are called super-operators. 

The quantum tunnelling super-operators $ {J}$ describe
the quantum `jumps' which instantaneously change the density matrix super-vector (we suppress the index $\alpha$ for a moment): 
\begin{equation}
 {J} |\rho ) = |\rho').
\end{equation}
The jump super-operators are written as a dyadic product of bra and ket supervectors:
\begin{equation}
 {J}  =  |  {J})  (\tilde{  {J}} |.
\end{equation}
Despite the notation, $ |  {J})  $ and $ (\tilde{  {J}} | $ are not
dual vectors related to each other via hermitian conjugation. 
Rather, they are formally defined as
\begin{equation}
 |\rho' ) = |  {J})  (\tilde{  {J}} |\rho),
 \end{equation}
 where  $ |\rho)$ is the density matrix before the quantum jump and 
  $ |\rho')$ contains the element  of the density matrix altered by the jump.
We will show, using particular examples, that  $ |  {J})  $ picks out the row from L and  $ (\tilde{  {J}} | $ 
picks out the column along with numerical value of the quantum jump operator.
Now, closely following  Brandes' derivation \cite{brandes08}, we are ready to simplify the expression (\ref{w}) for the tunelling time distribution:

\begin{eqnarray}
w(\tau) & =& \frac{ (I  |   {J}_{\text{md}})  ({  {\tilde J}_{\text{md}}} 
|e^{ {L}_{0}\tau} |  {J}_{\text{sm}})  ({  {\tilde J}_{\text{sm}} } | \overline\rho )}{(I |  {J}_{\text{sm}})  ({  {\tilde J}_{\text{sm}}} | \overline\rho )}
\nonumber
\\
&&= \frac{ (I  |   {J}_{\text{md}})  (  {\tilde J}_{\text{md}} |e^{ {L}_{0}\tau} |  {J}_{\text{sm}})  }{(I |  {J}_{\text{sm}}) }
\label{w2}
\end{eqnarray}

Since the quantum tunnelling operators change the diagonal elements of the density matrix (populations), then 
\begin{equation}
(I |  {J}_{\text{sm}}) = (I |  {J}_{\text{md}}) =1,
\end{equation}
the tunnelling time distribution becomes
\begin{equation}
w(\tau) = (  {\tilde J}_{\text{md}}
|e^{ {L}_{0}\tau} |  {J}_{\text{sm}}).
\label{w-final}
\end{equation}
As a result,  the steady state density matrix is not required for calculations of the tunnelling time distribution. The $w(\tau)$ are inherently real and positive quantities.  
The expression for the tunnelling time distribution  (\ref{w-final}) is the main equation that will be discussed in the following sections using a few particular examples.

\section{Calculations of the tunnelling time distribution for electron transport through a molecule with a single resonance molecular orbital}
Let us consider   a molecule attached to two macroscopic metal electrodes.
The total Hamiltonian is
\begin{equation}
 {H}= {H}_{S}+ {H}_{D}+ {H}_{M}+ V.
\end{equation}
The source and drain electrodes contain free electrons and are described by the following Hamiltonians:
\begin{equation}
  {H}_{S}=\sum_{l \sigma} \varepsilon_{l}a_{l\sigma }^{\dagger}a_{l \sigma} , \;\;\;\;\  {H}_{D}=\sum_{r \sigma }\varepsilon_{r }a_{r \sigma }^{\dagger}a_{r \sigma}.
 \end{equation}
 Here, $a^\dagger_{l\sigma/r \sigma}$   creates  an electron with spin $\sigma=\uparrow, \downarrow$ in the single-particle state $l/r$ of the source/drain electrode and $a_{l\sigma/r \sigma}$ is the corresponding electron  annihilation operator.
The molecule is described by a single spin degenerate molecular orbital with energy $\epsilon$
\begin{equation}
H_{M}=\epsilon\sum_{\sigma}a_{\sigma}^{\dagger}a_{\sigma},
\label{hm}
\end{equation}
where the
operator $a^\dag_\sigma (a_\sigma)$ creates (destroys) electron with spin $\sigma$ on the molecular level.
The tunnelling coupling between the molecule and electrodes is
 \begin{align}
V= t_S \sum_{l\sigma} (a_{l \sigma}^{\dagger}a_{\sigma }+  a^\dag_{\sigma }a_{l \sigma} ) + t_D \sum_{r\sigma}  (a_{r \sigma}^{\dagger}a_{\sigma }+ a^\dag_{\sigma }a_{r \sigma} ).
\end{align}

We assume that the electron repulsion is large in comparison with other relevant energy scales, so that the molecular orbital cannot 
accommodate  two electrons --  the molecule
can either be empty, or occupied by a spin up or a spin down electron.

Using the density matrix for the molecule we can introduce the following probabilities (molecular populations):
$\rho_{00} = \bra{0} \rho \ket{0} $ -- probability that the impurity is empty,
$\rho_{11} = \bra{\uparrow} \rho \ket{\uparrow}  + \bra{\downarrow} \rho \ket{\downarrow}  $ -- probability that the impurity is populated by one electron.
These probabilities are subject to normalisation condition
\begin{equation}
\rho_{00} + \rho_{11} =1.
\end{equation}
The relevant state vectors are applied to the Redfield master equation, and the matrix elements are collected to obtain the rate equation:
\begin{eqnarray*}
\left[\begin{array}{c}
\dot{\rho}_{00}\\
\dot{\rho}_{11}
\end{array}\right] & = & \left[\begin{array}{cc}
-2 T_{01} &  T_{10}\\
2 T_{01} & - T_{10}
\end{array}\right]\left[\begin{array}{c}
\rho_{00}\\
\rho_{11}
\end{array}\right].
\label{rate}
\end{eqnarray*}
The total rates are split into contributions from the source and the drain:
\begin{equation}
 T_{10} =  T^S_{10} +  T^D_{10},  \;\;\;\;  T_{01} =  T^S_{01} +  T^D_{01}.
\end{equation}
Here the partial rates are defined as
\begin{eqnarray}
 T^S_{01} &=&  \Gamma_S f_S,\;\;\;\;  T^D_{01} =  \Gamma_D f_D,
\\
 T^S_{10} &= &  \Gamma_S (1- f_S), \;\;\;\;  T^D_{10} =  \Gamma_D (1-f_D)
\end{eqnarray}
where $ \Gamma_{S/D} =2 \pi   |t_{S/D}|^2  \rho_{S/D}$ and $ \rho_{S/D}$ is the density of state for left/right electrode. 
$f_S$ and $f_D$ are Fermi occupation numbers for source and drain electrodes
\begin{equation}
f_S =[1+e^{(\epsilon-\mu_S)/T_S}]^{-1}, \;\;\;\; f_D= [1+e^{(\epsilon-\mu_D)/T_D}]^{-1}.
\end{equation}
Voltage bias is defined as the difference between the source and drain chemical potentials:
$V=\mu_S -\mu_D$.

Let us identify the quantum jump operators associated with electron tunnelling events. We write them in matrix form and as a  dyadic product of two supervectors.
There are two `intuitive' quantum jumps along the flow of the electrons: tunnelling from the source electrode to the  molecule
\begin{equation}
\label{jsm}
 {J}_{\text{sm}}  =  \left[\begin{array}{cc}
0 & 0 \\
2 T^S_{01} & 0
\end{array}\right]
= |  {J}_{\text{sm}})( {\tilde  J}_{\text{sm}}|
  =  \left[\begin{array}{c}
0\\
1
\end{array}\right]\left[\begin{array}{cc}
2 T_{01}^{S} & 0\end{array}\right]
\end{equation}
and from the molecule to the drain electrode
\begin{equation}
\label{jmd}
 {J}_{\text{md}}  = \left[\begin{array}{cc}
0 &  T_{10}^{D} \\
0 & 0
\end{array}\right]=  | { J}_{\text{md}})( {\tilde  J}_{\text{md}}|\\
  =  \left[\begin{array}{c}
1\\
0
\end{array}\right]\left[\begin{array}{cc}
0 &  T_{10}^{D}\end{array}\right].
\end{equation}
There are also  two `counterintuitive' quantum jumps in the Liouvillian superoperator where an electron attempts to jump against the voltage bias:
tunnelling from the drain electrode into the molecule
\begin{equation}
\label{jdm}
 {J}_{\text{dm}}  =  \left[\begin{array}{cc}
0 & 0 \\
2 T^D_{01} & 0
\end{array}\right]
= |  {J}_{\text{dm}})( { \tilde J}_{\text{dm}}|
  =  \left[\begin{array}{c}
0\\
1
\end{array}\right]\left[\begin{array}{cc}
2 T_{01}^{D} & 0\end{array}\right]
\end{equation}
and from the molecule back to the source electrode
\begin{equation}
\label{jms}
 {J}_{\text{ms}}  = \left[\begin{array}{cc}
0 &  T_{10}^{S} \\
0 & 0
\end{array}\right]=  | {J}_{\text{ms}})( { \tilde J}_{\text{ms}}|\\
  =  \left[\begin{array}{c}
1\\
0
\end{array}\right]\left[\begin{array}{cc}
0 &  T_{10}^{L}\end{array}\right].
\end{equation}

\begin{figure}[t!]
\begin{center}
\includegraphics[width=\columnwidth]{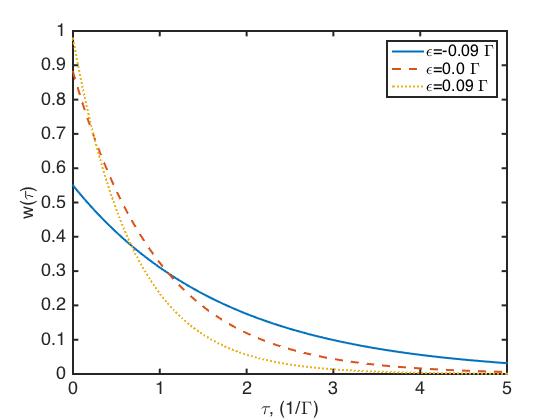}
\end{center}
\caption{ Tunnelling time distribution for different energies of the resonance molecular orbital. $T=0.05 \Gamma $, $\mu_S=0.1 \Gamma$,  $\mu_S=-0.1 \Gamma$, $\Gamma_L=\Gamma_R =\Gamma$. The Fermi energy $E_F$ is set to zero.}
\label{wtd}
\end{figure}

The Liouvillian without quantum jumps is diagonal
\begin{eqnarray*}
 {L}_{0} & = & \left[\begin{array}{cc}
-2 T_{01} & 0\\
0 & - T_{10}
\end{array}\right]
\end{eqnarray*}
and the tunnelling time distribution can be readily computed with the use of (\ref{w-final})
\begin{eqnarray}
w(\tau) &=& \left[\begin{array}{cc}
0 &  T_{10}^{D}\end{array}\right]  
\left[\begin{array}{cc}
e^{-2 T_{01} \tau} & 0\\
0 & e^{- T_{10} \tau}
\end{array}\right] \left[\begin{array}{c}
0\\
1
\end{array}\right] 
\nonumber
\\
 &=&  T_{10}^{D}e^{- T_{10}\tau}.
\end{eqnarray}
Substituting explicit expressions for the rates in terms of $\Gamma_{L/R}$ and Fermi occupation numbers, we get 
\begin{equation}
w(\tau) =  \Gamma_{D} (1- f_D) e^{-[ \Gamma_{S} (1- f_S) +  \Gamma_{D} (1- f_D)]\tau}.
\end{equation}
Fig.\ref{wtd} shows the tunnelling time distribution for electron transport through a molecular junction calculated for different positions of the resonance energy level relative to the electrodes' Fermi energy. The distribution is exponential, which indicates that electron tunnelling events described by a Markovian master equation are Poissonian statistical processes.
The exponential distribution shown in Fig.\ref{wtd} is qualitatively similar to the distribution obtained by Carmichael, Singh, Vyas, and Rice in their 1989 paper \cite{PhysRevA.39.1200}. They derived an exponential distribution for the waiting time between photon counts from a coherently driven two level atom. 

\begin{figure}[t!]
\begin{center}
\includegraphics[width=\columnwidth]{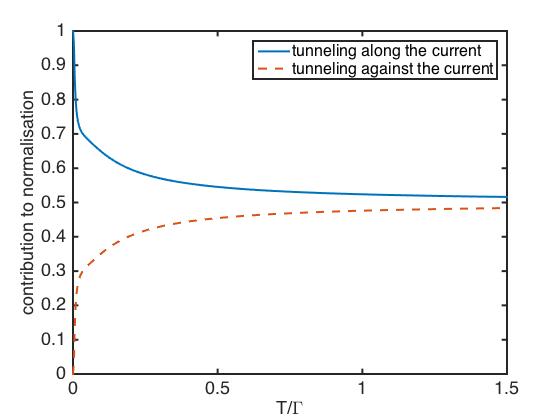}
\end{center}
\caption{Overall contribution to normalisation from different tunnelling quantum jumps as a function of temperature: solid line  - jumps along the current flow eq.(\ref{norm1}), dashed line - jumps against the current flow eq.(\ref{norm2}). Parameters used in calculations:  $\epsilon=-0.09\Gamma$, $\mu_S=0.1 \Gamma$,  $\mu_S=-0.1 \Gamma$, $\Gamma_L=\Gamma_R =\Gamma$. The Fermi energy $E_F$ is set to zero.}
\label{normalisation}
\end{figure}

The contribution to the normalisation from the tunnelling time distribution  (in other words, the contribution from the distribution describing an electron transferred from the source to the molecule and then from the molecule to the drain)
is
\begin{equation}
\int_0^\infty d\tau  w(\tau) = \frac{  \Gamma_{D} (1- f_D)}{  \Gamma_{S} (1- f_S) +  \Gamma_{D} (1- f_D)}.
\label{norm1}
\end{equation}
The rest of the normalisation comes from the backscattering processes where an electron first tunnels into the molecule and then tunnels back to the source electrode
\begin{equation}
\int_0^\infty d\tau  (  {\tilde J}_{\text{ms}}
|e^{ {L}_{0}\tau} |  {J}_{\text{sm}}) = \frac{  \Gamma_{S} (1- f_S)}{  \Gamma_{S} (1- f_S) +  \Gamma_{D} (1- f_D)}.
\label{norm2}
\end{equation}
The last tunnelling event  $(  {\tilde J}_{\text{dm}}
|e^{ {L}_{0}\tau} |  {J}_{\text{sm}}) $ is zero due to the Pauli principle. The contributions to the normalisation from different secondary quantum jumps can be used to quantify the relative likelihood of extreme events in electron transport - tunnelling against the average electric current flow. Fig.\ref{normalisation} shows the contributions to the overall tunnelling probability of the tunnelling time and backscattering distributions as a function of temperature. At low temperature, tunnelling from the source to the drain dominates the electron transport. However, as the temperature increases the number of electrons jumping against the voltage basis increases and finally both processes asymptote to the same contribution in the high temperature limit.

The average tunnelling time and standard
deviation are the first and second moments of the tunnelling time distribution respectively:
\begin{eqnarray*}
\langle\tau\rangle & = & \int_{0}^{\infty}\tau\,w(\tau)d\tau= \frac{  \Gamma_{D} (1- f_D)}{[  \Gamma_{S} (1- f_S) +  \Gamma_{D} (1- f_D)]^{2}},\\
\sigma & = & \sqrt{\langle\tau^{2}\rangle-\langle\tau\rangle^{2}}=\frac{\sqrt{ \Gamma^2_{S} (1- f_S)^2 + \Gamma^2_{D} (1- f_D)^2 }}{ [  \Gamma_{S} (1- f_S) +  \Gamma_{D} (1- f_D)]^{2}}.
\end{eqnarray*}

Let us analyse the tunnelling time distribution's dependence on junction temperature. In the resonance transport regime the energy level is located between the source and drain Fermi energies, $\mu_{S}<\epsilon<\mu_{D}$, and the voltage bias is small.  When $T\rightarrow 0$, then $f_L(\epsilon)=1$ and $f_R(\epsilon)=0$, and the tunnelling time distribution becomes
\begin{equation}
w(\tau) =  \Gamma_{D}  e^{-  \Gamma_{D} \tau}.
\end{equation}
We see that at zero temperature, the tunnelling time distribution is determined solely by the coupling strength between the molecule and the drain electrode.

In the high temperature limit, $f_S=f_D=1/2$, the tunnelling time distribution is
\begin{equation}
w(\tau) = \frac{1}{2}  \Gamma_{D}  e^{-[ \Gamma_{S} +  \Gamma_{D} ]\tau/2}.
\end{equation}
In this case, the exponential decay depends symmetrically on the couplings to the source and drain electrodes, as the high temperature eliminates the electric current and the electrons have an equal probability of tunnelling to either the source or drain electrode from the molecule.

\begin{figure}[t!]
\begin{center}
\includegraphics[width=\columnwidth]{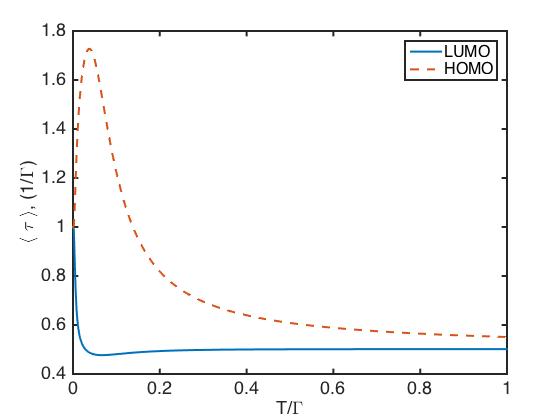}
\end{center}
\caption{Average waiting time for HOMO  resonance energy level which corresponds to  ($\epsilon=-0.09\Gamma$) and LUMO  resonance energy level ( $\epsilon=0.09\Gamma$) as a function of temperature. The parameters used in calculations: $\mu_S=0.1 \Gamma$,  $\mu_S=-0.1 \Gamma$, $\Gamma_S=\Gamma_D =\Gamma$. The Fermi energy $E_F$ is set to zero.}
\label{homo-lumo}
\end{figure}

The tunnelling time distribution provides additional information about electron transport mechanisms that is not available in current-voltage characteristics. 
For example, it is very difficult to unambiguously deduce from molecular conductance measurements  if a molecular junction is p-type or n-type, or in other words, if the position of the Fermi level $E_F$ of the metal electrode is closer to the highest occupied molecular orbital (HOMO) or to the lowest unoccupied molecular orbital (LUMO) \cite{reddy07}. Although directly measuring the Seebeck coefficient by applying a temperature gradient along the molecular junction distinguishes  electron and hole conductivities \cite{reddy07}, maintaining the temperature gradient as opposed to maintaining voltage bias is much more experimentally challenging.

Fig.\ref{homo-lumo} shows that the tunnelling times of
 p-type and  n-type 
molecular junctions have distinctly different temperature dependences.
In our calculations we assumed that the molecule is symmetrically coupled to the source and drain electrodes such that:
\begin{equation}
\Gamma_L = \Gamma_R = \Gamma.
\end{equation}
The energy level of the HOMO orbital is  $\epsilon/\Gamma=-0.09 $. At $T=0$ the average waiting time is $1/\Gamma$. It rises to a peak at $T/\Gamma=0.05$, then decreases asymptotatically. Qualitatively opposite behaviour is observed for the LUMO energy level.

Let us understand the physical mechanisms responsible for these very different temperature dependences.
 At zero temperature,  $f_{S}=1$  and  $f_{D}=0$.
Therefore, the average waiting time at $T=0$ is $1/\Gamma$  for electron transport through both the HOMO and LUMO channels. As the temperature
increases the occupation number $f_{S}$ decreases, as there
is now a non-zero probability that electron states with energies higher
than $\mu_{S}$ are filled, and $f_{D}$ increases, as there
is now a non-zero probability that electron states with energies higher
than $\mu_{D}$ are filled. 
For the HOMO transport channel,
 $\epsilon$ is much closer to $\mu_{D}$
than $\mu_{S}$, so that small increases in temperature (of the order of $\epsilon-\mu_{D}$)
cause a much greater increase in $f_{D}$ than they cause a decrease
in $f_{S}$. Consequently, when temperature is of the order of $\epsilon-\mu_{D}$
the average waiting time is:
\[
\langle\tau\rangle = \frac{1-f_{D}}{(1-f_{D})^{2}}\frac{1}{\Gamma}.
\]
The average waiting time increases in this regime, which
is shown in the peak of the HOMO plot. However, as temperature increases
beyond the scale of $\epsilon-\mu_{D}$ the contribution from $f_{S}$
becomes non-negligible in comparison to the contribution from $f_{D}$.
The average waiting time becomes:
\begin{eqnarray}
\langle\tau\rangle & = & \frac{1-f_{D}}{(2-f_{S}-f_{D})^{2}} \frac{1}{\Gamma},\label{eq:HOMO_decrease}
\end{eqnarray}
so as $f_{S}$ decreases and $f_{D}$ increases the average waiting
time decreases. If the temperature is much greater than the energy
scales of the system, such that $T\gg\epsilon-\mu_{R}$, then the
average waiting time reduces to $ \langle\tau\rangle =\Gamma/2$.
It is the  common asymptote of both plots in Fig.\ref{homo-lumo}.

Now we consider the LUMO case, where $\epsilon$ is close to $\mu_{L}$.
Again, at $T=0$ the occupation numbers $f_{S}$ and $f_{D}$ are
unity and zero respectively and $\langle\tau\rangle=1$. Now however,
a small increase in temperature causes the average waiting time to
decrease. When the temperature is small the decrease in
$f_{S}$ is much greater than the increase in $f_{D}$ and the average
waiting time reduces to:
\begin{eqnarray*}
\langle\tau\rangle & = & \frac{1}{(2-f_{S})^{2}} \frac{1}{\Gamma},
\end{eqnarray*}
which is less than $1$. However, as the temperature increases
further the contributions from $f_{S}$ and $f_{D}$ become comparable
in size and we have the same situation as (\ref{eq:HOMO_decrease}),
except in this case $f_{S}$ is smaller at comparable temperatures
and $f_{D}$ is larger. Again, as $T\rightarrow\infty$ $\langle\tau\rangle\rightarrow =\Gamma/2$.

\section{conclusions}

We have considered electron transport through a molecular junction as a series of quantum events (electron tunnellings) separated by random time intervals.
We focus on the conditional probability density
that given an electron has tunnelled into the molecule from the source electrode
at time $t$, it tunnels out to the drain at time $t+\tau$ where $\tau$ is an inherently stochastic quantity. The tunnelling time distribution is defined as 
 this conditional probability density.
We introduced a derivation for the tunnelling time distribution and
defined it in terms of quantum tunnelling events.
Using superoperator formalism and the Markovian master equation for the time-evolution of the reduced density matrix of the molecule, we derived a computationally convenient but yet general expression 
for the tunnelling time distribution.

The theory was applied to electron transport through a model molecular junction with a single resonance molecular orbital. We obtained compact and useful for qualitative physical analysis analytic expressions for the tunnelling time 
distribution, the average tunnelling time and the tunnelling time dispersion. 
The tunnelling time distribution is exponential, which indicates that quantum tunnelling driven by a Markovian dissipative environment is a Poissonian statistical process.
The tunnelling time distribution is used to identify extreme statistical quantum events, such as electrons tunnelling against the average current flow.
The role of these extreme events in molecular conductivity is analysed and quantified by partitioning the overall normalisation of the secondary quantum tunnelling distribution between different quantum events. The average tunnelling times of HOMO and LUMO channels have completely different temperature dependences, which implies that 
tunnelling time distributions can be used to extract information about the alignment of molecular orbital relative to the electrode Fermi energy.


\appendix

\section{Proof of some useful identities}
\subsection{$ \text{Tr}[  {L}  A]=0$ for arbitrary linear operator $A$. }
The Markovian Liouville equation (\ref{liouville}) has the formal solution
\begin{equation}
 \rho(t) = e^{ {L} t} \rho(0).
\end{equation}
The density matrix should remain normalised during the time evolution
\begin{equation}
1= \text{Tr}[ \rho(t)].
\end{equation}
Differentiating this equation with respect to time yields
\begin{equation}
0= \text{Tr}[  {L}  \rho(t)].
\label{0}
\end{equation}
Since eq.(\ref{0}) is valid for any time $t$, it is also valid for $t=0$:
\begin{equation}
0= \text{Tr}[  {L}  \rho(0)].
\label{00}
\end{equation}
The initial condition is an arbitrary (Hermitian, normalized $\text{Tr}[ \rho(0)] =1$) operator.
Being valid for an arbitrary hermitian operator, 
Eq.(\ref{00}) is  also satisfied for any linear operator $A$  (since any linear operator $A$ can be written as $A= B+iC$, where $B$ and  $C$ are Hermitian), therefore
\begin{equation}
 \text{Tr}[  {L}  A]=0.
\end{equation}

\subsection{$(I |  {J}_{\text{sm}}) = (I |  {J}_{\text{md}}) =1$}

In super-operator form the density matrix becomes a vector $|\rho\rangle=[\rho_{00},\rho_{11},..,\rho_{NN},\rho_{01},...,\rho_{NN-1}]^{T}$
so that the first $N$ elements are the diagonals of the density matrix.
Therefore, we define a row vector $(I|=[1,1,...,1,0,...,0]$, with
ones for the first $N$ elements and zeroes elsewhere so that the trace
operation can be written as:

\begin{eqnarray*}
\text{Tr}[A\rho] & = & (I|A\rho).
\end{eqnarray*}

The jump operators $ {J}_{\text{sm}}$ and $ {J}_{md}$ can be
written as a dyadic product:

\begin{eqnarray*}
 {J}_{\text{sm}} & = & | {J}_{sm})( {\tilde J}_{sm}|,
\end{eqnarray*}

and similarly for $ {J}_{\text{dm}}$. Here, the column vector $| {J}_{\text{sm}})$
picks out the row of the jump operation, so it has a value of one
in the corresponding row and zeroes elsewhere. Since the tunnelling
events are concerned with transitions from one pure state to another,
the row of the jump operation is always in the first $N$ elements.
As a result, the inner product between the trace operation and the
jump operation is always unity:

\begin{eqnarray}
(I| {J}_{\text{sm}}) & = & [1,1,...,1,0,...,0]\cdot[0,1,0,...,0]^{T}\\
 & = & 1\\
(I| {J}_{\text{md}}) & = & [1,1,...,1,0,...,0]\cdot[1,0,0,...,0]^{T}\\
 & = & 1
\end{eqnarray}

\end{document}